\documentclass{aa}
\usepackage{txfonts}
\usepackage{epsfig,graphics,longtable,amssymb,verbatim}
\usepackage{natbib}
\bibpunct{(}{)}{;}{a}{}{,} %voor A&A stijl

\begin{document}

\title{Radio observations of candidate magnetic O stars}

\author{R.S. Schnerr\inst{1}
      \and K.L.J. Rygl\inst{1}
      \and A.J. van der Horst\inst{1}
      \and T.A. Oosterloo\inst{2}
      \and J.C.A. Miller-Jones\inst{1}
      \and H.F. Henrichs\inst{1}
      \and T.A.Th.~Spoelstra\inst{2}
      \and A.R. Foley\inst{2}
          }

\institute{Astronomical Institute ``Anton Pannekoek'', University of Amsterdam, Kruislaan 403, 1098 SJ Amsterdam, Netherlands
  \and     ASTRON, 7991 PD, Dwingeloo, Netherlands
}

\offprints{R.S. Schnerr,
\email{rschnerr@science.uva.nl}}

\date{Received date / Accepted date}


\abstract
{Some O stars are suspected to have to have (weak) magnetic fields because of the observed cyclical variability in
their UV wind-lines. However, direct detections of these magnetic fields using optical spectropolarimetry have proven to be very difficult.}
{Non-thermal radio emission in these objects would most likely be due to synchrotron radiation. As a magnetic field is required for the production of synchrotron radiation, this would be strong evidence for the presence of a magnetic field. Such non-thermal emission has already been observed from the strongly magnetic Ap/Bp stars.}
{We have performed 6 \& 21 cm observations using the WSRT and use these, in combination with archival VLA data at 3.6 cm and results from the literature, to study the radio emission of 5 selected candidate magnetic O stars.}
{Out of our five targets, we have detected three: $\xi$ Per, which shows a non-thermal radio spectrum, and $\alpha$ Cam and $\lambda$ Cep, which show no evidence of a non-thermal spectrum. In general we find that the  observed free-free (thermal) flux of the stellar wind is lower than expected. This is in agreement with recent findings that the mass-loss rates from O stars as derived from the H$\alpha$ line are overestimated because of clumping in the inner part of the stellar wind.
}
{}

\authorrunning{R.S.\ Schnerr et al.}
\maketitle

\section{Introduction}
All O-type stars have strong, line-driven winds. They usually have a thermal radio spectrum due to the free-free emission from the ionised stellar wind \citep{abbott:1980,bieging:1989,scuderi:1998}.
This spectrum can be calculated using:

\begin{eqnarray}
\label{eq:thermal}
S_\nu &=& 7.26 \left( \frac{\nu}{10\,\mathrm{GHz}} \right)^{0.6} \left( \frac{T_\mathrm{e}}{10^{4}\,\mathrm{K}} \right)^{0.1}
          \left( \frac{\dot{M}}{10^{-6}\,\mathrm{M_\odot \, yr^{-1}}} \right)^{4/3} \nonumber\\
      & & \left(\frac{\mu_\mathrm{e} \mathrm{v}_\infty}{100\, \mathrm{km\,s^{-1}}} \right)^{-4/3} \left(\frac{D}{\mathrm{kpc}} \right)^{-2}  \mathrm{mJy},
\end{eqnarray}
\citep{wright:1975,panagia:1975,scuderi:1998}, where D is the distance to the star, $\dot{M}$ is the mass-loss rate, $T_\mathrm{e}$ the electron temperature, $\mu_\mathrm{e}$ the mean atomic weight per electron, $\mathrm{v}_{\infty}$ the terminal wind velocity and $\nu$ the observing frequency.

However, about 30\% of the O stars are found to show non-thermal radio emission \citep[see, e.g.,][]{bieging:1989,drake:1990,scuderi:1998,benaglia:2001}. This is characterised by a flatter than thermal spectrum, i.e.\ defined as $\alpha < 0.6$, with $S_\nu \propto \nu^\alpha$.
\citet{white:1985} proposed that synchrotron radiation from the rapidly moving electrons of the wind in a
stellar magnetic field could also contribute to the radio emission. This was confirmed by the discovery of non-thermal
radio emission in the magnetic Ap/Bp stars \citep[][see also \citealt{cassinelli:1984}, who report non-thermal emission found in $\sigma$ Ori E by Churchwell]{drake:1987}. These stars have strong (of the order of kilogauss), dipole-like magnetic fields. Among O stars only two magnetic stars are
known: \object{$\theta^1$ Ori C} \citep{donati:2002} and \object{HD 191612} \citep{donati:2006}. Nevertheless, strong indirect evidence exists that many 
O stars should have magnetic fields \citep[e.g.][]{henrichs:2005}. One of the main arguments why many O stars are thought to have (weak) 
magnetic fields is that their winds show cyclic behaviour on a rotational timescale, which is typically a few days \citep[see][ for a review]{fullerton:2003}. 
The lack of magnetic field detections is most likely related to the fact that
direct measurements of magnetic fields in O stars are extremely difficult, because of the very few available
spectral lines in the optical region.
The usual method to measure magnetic fields is to determine the magnetic Zeeman splitting of magnetically sensitive lines with optical spectropolarimetry. The sensitivity of this method decreases towards earlier spectral types, as these stars have fewer spectral lines in the optical region.

The detection of non-thermal radio emission from O stars with such cyclic variability would be strong evidence that magnetic fields are indeed present in these stars. We have selected five candidate magnetic O stars that have been studied extensively in the ultraviolet (UV) in order to search for evidence of non-thermal radio emission.

\begin{table*}[!t]
\begin{center}
\caption{Stellar parameters of the selected O stars. Spectral types are from \citet{walborn:1973,walborn:1976}. Hipparcos parallaxes were taken from \citet{perryman:1997}. All other parameters were taken from the ``preferred solution'' (some stars have more than one possible set of parameters that fits the observations) of \citet{markova:2004}, except for 10 Lac, for which we used \citet{mokiem:2005}.
We show both the distance from Hipparcos and the distance used for the spectral modeling. As this latter value is used to determine the mass-loss rates, this distance was also used for predicting the thermal radio flux. For the preferred solutions of \citet{markova:2004}, we have scaled the distance from the original solution using the absolute magnitudes, as the reddening is assumed to be the same for both solutions.}
\label{targets}
\begin{tabular}{l|lllll}
\hline\hline
                                             & \object{$\xi$ Per}   & \object{$\alpha$ Cam} & \object{15 Mon}	& \object{$\lambda$ Cep} & \object{10 Lac}	      \\
\hline	     
HD number                            	     & 24912		    & 30614		    & 47839		& 210839		& 214680		       \\
Association/Runaway                          & Runaway              & Runaway               & Mon OB1           & Runaway               & Lac OB1 \\
Spectral type                        	     & O7.5III(n)((f))$^\mathrm{a}$  & O9.5 Ia	    & O7 V ((f))	& O6 I (n)fp		& O9 V  		       \\	
Parallax  (mas)                      	     & $1.84\pm0.70$	    & $0.47\pm0.60$	    & $3.09\pm0.53$	& $1.98\pm0.46$ 	& $3.08\pm0.62$ 	       \\
Hip. distance  (pc)                  	     & $540^{+330}_{-150}$  & $>$440		    & $323^{+67}_{-47}$ & $510^{+150}_{-100}$	& $320^{+90}_{-50}$	       \\     
Spectral mod. distance (pc)              	     & 895		    & 765		    & 710		& 1089  		& 320\\
Mass ($\mathrm{M}_\odot$)                    & 52                   & 22		    & 25		& 58			& 27     		       \\  
Radius ($\mathrm{R}_\odot$)           	     & 25.2		    & 19.6		    & 9.9		& 23			& 8.3           	       \\  
$T_\mathrm{eff}$  ($10^3$ K)           	     & 34.0      	    & 31.0		    & 37.5		& 36.2  		& 36.0	       \\
Luminosity ($10^5 \mathrm{L}_\odot$)         & 7.6                  & 3.2		    & 1.7		& 8.1			& 1.0	       \\ 
Terminal velocity (km s$^{-1}$)              & 2400       	    & 1550		    & 2200		& 2200  		& 1140		    \\
Mass-loss ($10^{-6}$M$_\odot$ y$^{-1}$)      & $4.0 \pm 1.0$        & $2.9\pm0.9$	    & $1.2\pm0.3$	& $7.7\pm2.3$    	& $6.1^{+8.8}_{-5.5} \times 10^{-2}$\\
\hline
\end{tabular}
\end{center}
\begin{list}{}{}
\item[$^{\mathrm{a}}$] This is the spectral type given by \citet{walborn:1973}, however, a spectral type of O7.5I(n)((f)) was adopted by \citet{markova:2004}.
\end{list}
\end{table*}

\section{Observations \& data reduction}
\label{obs&data}
For this study five targets have been selected from the 10 O stars listed by \citet{kaper:1996}, which are the brightest and best studied O stars in the UV with the  International Ultraviolet Explorer (IUE) satellite. All these targets show extensive stellar wind variability, some with well studied cyclic behaviour. The final selection was made on the criteria that the star should be observable with the Westerbork
Synthesis Radio Telescope (WSRT) and that the star should have been previously detected in the radio region ($\alpha$~Cam, 15~Mon and $\lambda$~Cep; for $\xi$~Per archive observations from the Very Large Array --VLA-- were available). We added 10 Lac because of its brightness 
and its rich UV history. The stellar parameters of these stars are listed in Table~\ref{targets}.

The radio observations of our targets selected from the literature have been summarised in Table~\ref{litdet}. The detection of a $42\pm5$ mJy source near the optical position of $\xi$ Per by \citet{bohnenstengel:1976} at 11 cm (2.7 GHz) is discussed in Sect.~\ref{results:xiper}.

To complement these measurements, and in order to determine the spectral slopes, we have used WSRT (6 and 21 cm) and VLA (3.6 cm) observations.

\begin{table}[htb]
\caption{Radio detections as a function of wavelength of our targets reported in the literature. When several measurements are available a weighted average is shown; upper limits are 3$\sigma$. Fluxes are from [1] \citet{abbott:1980}, [2] \citet{bieging:1989}, [3] \citet{drake:1990}, [4] \citet{lamers:1993} and [5] \citet{scuderi:1998}.}
\label{litdet}
\begin{center}
\begin{tabular}{l @{~~} c @{~~} c @{~~} c @{~~} c}
\hline
\hline
Star          & \multicolumn{3}{c}{Flux (mJy)} \\
              &  2 cm            & 3.6 cm          & 6 cm            & References\\
\hline
$\alpha$ Cam  &  $0.65\pm0.13$   & $0.44\pm0.04$   & $0.29 \pm 0.04$ & 2,5\\
15 Mon        &  $<$0.4	         &		   & $0.40 \pm 0.13$ & 3\\
              &                  &                 & $<$0.33 \& $<$0.18   & 2,3\\
$\lambda$ Cep & 		 & $0.38 \pm 0.03$ & $0.40 \pm 0.25$ & 1,4\\
\hline
\end{tabular}
\end{center}
\end{table}

We performed observations for all five selected O stars at 21 cm (1.4 GHz) with the WSRT during the period from September to November 2005. In addition, 10 Lac was
observed at 6 cm (4.9 GHz). All observations consisted of 12 h integrations in the Maxi-Short configuration, done in continuum mode with a bandwidth of 8$\times$20 MHz.
Gain and phase calibrations were done using the calibrator 3C286, except the observation of $\alpha$ Cam which was calibrated with 3C48.

Earlier observations of $\xi$ Per were performed in 1995 at 6 cm (6 observations in May and June, total of 26 h)
, and 21 cm (5 observations in June, July and August, total of 39 hours). Due to the lower sensitivity of the WSRT at that time, the observations at each frequency were all combined. The 21 cm observations were calibrated using 3C48 and the 6 cm observations were calibrated using 3C48, 3C147 and 3C286.

The reduction of the WSRT data was done using the {\sc miriad} software package. 

From the VLA archive we used an X-band (3.6 cm, 8.5 GHz) continuum observation taken on 11 Jan 1999 (program ID AS644-x) of $\xi$ Per. This 0.68h observation was taken in C configuration with a bandwidth of 50 MHz. The data were reduced with {\sc aips}, using 3C48 as a primary and B0411+341 as a secondary calibrator.

\subsection{Distances and mass-loss rates}
As massive stars are relatively far away, their distances as determined by the Hipparcos satellite suffer from systematic errors and and may be underestimated \citep[e.g.][]{schroder:2004}.
Distances may also be obtained using the stellar properties derived from spectral modeling. The spectral modeling distance for a star is generally derived from the relation between the spectral type and luminosity, possibly making use of other stars in the same cluster. This distance is used to determine the absolute magnitude of the star and this constrains the stellar radius. As the mass-loss rates determined in this way depend on the radius, we have adopted the distances used in the spectral modeling to calculate the predicted radio fluxes. This ensures that our mass-loss rates and distances are consistent (see also Sect.~\ref{clumping}).

\section{Results}
\begin{figure*}[phtb]
\begin{center}
\includegraphics[width=\linewidth,height=0.5\linewidth]{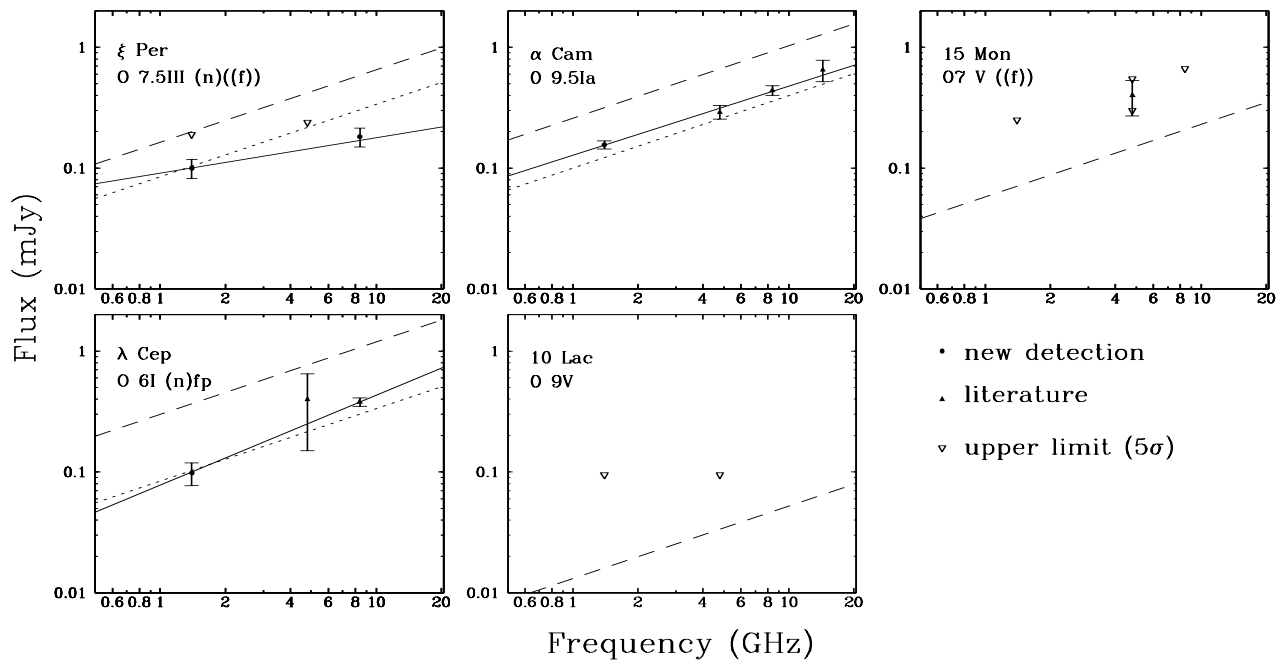}
\caption[]{Radio spectra of the 5 selected targets. Shown are the new results, results taken from the literature, upper limits, the predicted thermal flux (Eq.\ \ref{eq:thermal}, dashed line), the predicted thermal flux accounting for clumping \citep[][dotted line]{puls:2006}, and a fit to the observations (solid line). Distances used for estimating the thermal flux are the same as those used for the determination of the mass-loss rates using H$\alpha$.}
\label{spectra}
\end{center}
\end{figure*}

We detected three of our five selected targets (see Table~\ref{results_table}). For $\xi$ Per this was the first detection in the radio; all three stars have now been detected for the first time at 21 cm. In general the flux was found to be lower than the predicted thermal flux using Eq.~\ref{eq:thermal} and the stellar parameters from Table~\ref{targets} \citep[assuming typical values of $\mu_e=1.3$ and $T_\mathrm{e}=0.85$ $T_\mathrm{eff}$;][]{scuderi:1998}. Both the predicted thermal flux and a power law fit to the observations are shown in Fig.~\ref{spectra}.

We now present the results per source in order of RA:

\begin{table*}[htbp]
\begin{center}
\caption[]{Results of our new WSRT and archival VLA observations. The upper limits shown are 5$\sigma$ upper limits.}
\label{results_table}
\begin{tabular}{rllcccll}
\hline
\hline
\multicolumn{1}{c}{HD} & star & date         & freq.   & $\lambda$   & flux       & spectral	  & array \\
number                 & name & (d/m/y)      & (GHz)   & (cm)        & $\mu$Jy    & index $\alpha$&	  \\
\hline
 24912   & $\xi$ Per          & May/Jun/95   & 4.9     &  6          & $\leq$240  &               & WSRT  \\
         &                    & Jun/Aug/95   & 1.4     & 21          & $\leq$190  &               & WSRT  \\
         &                    & 11/Jan/99    & 8.4     &  3.6        & 169$\pm$30 & 0.29$\pm$0.14 & VLA   \\
         &                    & 28/Nov/05    & 1.4     & 21          & 100$\pm$18 &		  & WSRT  \\
 30614   & $\alpha$ Cam       & 09/Oct/05    & 1.4     & 21          & 156$\pm$12 & 0.57$\pm$0.06 & WSRT  \\
 47839   & 15 Mon             & 13/Oct/05    & 1.4     & 21          & $\leq$250  &		  & WSRT  \\
210839   & $\lambda$ Cep      & 03/Oct/05    & 1.4     & 21          & 98$\pm$21  & 0.74$\pm$0.11 & WSRT  \\
214680   & 10 Lac             & 22/Sep/05    & 4.9     &  6          & $\leq$95   &		  & WSRT  \\
         &                    & 21/Sep/05    & 1.4     & 21          & $\leq$95   &		  & WSRT  \\
\hline
\hline
\end{tabular}
\end{center}
\end{table*}

\subsection{$\xi$ Per}
\label{results:xiper}
In the 1995 WSRT observations, $\xi$ Per was not detected at 6 and 21 cm. However, it was detected in the higher S/N 3.6 cm (VLA, Jan 1999) and 21 cm (WSRT, Nov 2005) observations. The flux was found to be lower than the predicted thermal flux by a factor of $\sim$2 (21 cm) to $\sim$3.5 (3.6 cm). The spectrum has a spectral index of $\alpha=0.29\pm0.14$, which is lower than the thermal value of 0.6. This is evidence for the presence of a non-thermal contribution to the observed flux.

\citet{puls:2006} found an upper limit for $\xi$ Per of 120 $\mu$Jy (3$\sigma$) from VLA observations on March 9, 2004. As this limit is not consistent with our detection at this wavelength, this might be an indication of variability. To check this, we retrieved the observations from the VLA archive. At the position of $\xi$ Per, we measured a flux density of $154\pm39$ $\mu$Jy, which, we agree, is not a reliable detection of the source (3.9$\sigma$). However, it is consistent with our detection of the source at $169\pm30$ $\mu$Jy in 1999.

\begin{figure}[htb]
\begin{center}
\includegraphics[width=\linewidth]{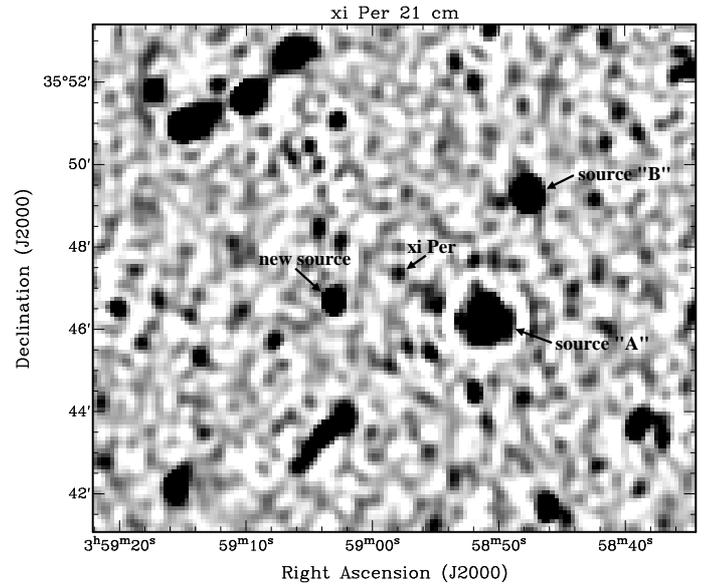}
\caption[]{Our 21 cm $\xi$ Per observation of 2005. The sources A and B of \citet{bohnenstengel:1976} are indicated, as are $\xi$ Per and a new source that they did not detect (see Section \ref{results:xiper}).}
\label{bohnenstengel}
\end{center}
\end{figure}

\citet{bohnenstengel:1976} detected a source of $42\pm5$ mJy at 2.695 GHz near the optical position of $\xi$ Per. They concluded that this component is either due to an extended ($\sim$2$\arcmin$) thermal source of about 10 mJy, or to blending of their components A and B (see Fig.~\ref{bohnenstengel}), in which case they claim that component B has to have a very flat spectrum. As an interferometer such as the WSRT is not very sensitive to extended structures, we cannot exclude the presence of an extended source. We find that component B has a spectral index of $\alpha$$\approx$$-1.1$. It is detected at 5.8$\pm$0.2 mJy at 21 cm, its 6 cm flux is 1.5$\pm$0.1 mJy and at 3.6 cm it shows extended structure (10$\arcsec$x2$\arcsec$) but has very low flux density (peak of $\sim$0.2 mJy/beam). At 3.6 and 6 cm, component A can be resolved into two components with a separation of $12\pm1 \arcsec$. The western component has a flux of $6.1\pm1.1$ (6 cm) and $2.6\pm0.5$ mJy (3.6 cm) and the eastern component $7.1\pm0.9$ (6 cm) and $3.0\pm0.5$ mJy (3.6 cm). In addition we found a third source at the position [$\alpha$(2000)=03h\,59m\,03s, $\delta$(2000)=+35$^\circ$46$\arcmin$42$\arcsec$], which is not detected at 6 cm ($\leq0.24$ mJy) but is detected at 0.9$\pm$0.2 mJy at 21 cm.

Given the low resolution of the Effelsberg telescope ($\sim$5$\arcmin$ at 11 cm) compared to the WSRT ($\sim$15$\arcsec$ at 21 cm) and VLA ($\sim$2$\arcsec$ at 3.6 cm) observations, we conclude that it is very likely that the source found by \citet{bohnenstengel:1976} is due to blending of all these components.

\subsection{$\alpha$ Cam}
This star shows a thermal spectrum over the entire range from 2 to 21 cm. The flux is found to be lower than the predicted thermal flux by a factor of $\sim$2.

\subsection{15 Mon}
\citet{drake:1990} found a 3$\sigma$ upper limit of 400 $\mu$Jy on 24 Jan 1987 for this source at 2 cm. At 6 cm \citet{bieging:1989} reported a 3$\sigma$ upper limit of 330 $\mu$Jy, and \citet{drake:1990} found an upper limit of 180 $\mu$Jy on 22 Feb 1986. However, one year later (24 Jan 1987) \citet{drake:1990} detected the source at the same wavelength at $400 \pm 130$ $\mu$Jy. This suggests that the radio flux of 15 Mon is possibly variable.

Due to the unfavourable position, for an E-W array, of 15 Mon on the sky (close to the equator), our 5$\sigma$-upper limit of 250 $\mu$Jy at 21 cm is relatively high compared to the expected thermal noise for a 12h run.

\subsection{$\lambda$ Cep}
Our detection of $\lambda$ Cep at 21 cm allows for an accurate determination of the spectral index. We find that $\lambda$ Cep has an approximately thermal spectrum, with a spectral index of $\alpha=0.74\pm0.11$. The flux is, however, a factor of $\sim$3--3.5 lower than predicted.

\subsection{10 Lac}
We have not detected 10 Lac, with a 5$\sigma$ upper limits of 95 $\mu$Jy at both 21 and 6 cm. This is in agreement with the expected thermal flux based on the determination of the mass-loss rate by \citet{mokiem:2005} of $6.1^{+8.8}_{-5.5}\times 10^{-8} \mathrm{M}_\odot \mathrm{y}^{-1}$. 
Our 6 cm upper limit constrains the mass-loss rate to be lower than $1.3 \times 10^{-7} \mathrm{M}_\odot \mathrm{y}^{-1}$.

\section{The effects of clumping in stellar winds}
\label{clumping}
As recently discussed by \citet{fullerton:2006} and \citet{puls:2006}, it is generally found that mass-loss rates determined from H$\alpha$ are higher by a factor of $\sim$3-8 than those derived from radio observations. This is thought to be due to enhanced clumping in the inner part of the wind where the H$\alpha$ emission originates, compared to the outer part of the wind from which we receive the observed radio emission. As both the H$\alpha$ and radio emission are proportional to the density squared, clumping in the wind results in an overestimate of the mass-loss rate. Since we use mass-loss rates derived from H$\alpha$ to predict the thermal radio fluxes, the fact that the observed fluxes are lower than our predictions is consistent with stronger clumping in the inner part of the wind compared to the outer part.

We have used the same distances as assumed in the H$\alpha$ modeling. For the H$\alpha$ modeling the distance (or absolute magnitude) is used to estimate the stellar radius, resulting in $R \propto D$. In these models the mass-loss rate approximately scales with $R$ as $\dot{M} \propto R^{3/2}$, which gives $\dot{M} \propto D^{3/2}$. Since the predicted radio flux scales as $S_\nu \propto \dot{M}^{4/3} D^{-2}$ (Eq.~\ref{eq:thermal}), one finds that the predicted radio flux is approximately independent of the distance.

For $\alpha$ Cam and $\lambda$ Cep, our results are in good agreement with the mass-loss rates determined by \citet{puls:2006}, and for $\xi$ Per, the flux observed at 3.6 cm is in agreement with the upper limits of both possible solutions quoted.

\section{Conclusions}
\label{conclusions}

We have detected three candidate magnetic O stars at radio wavelengths. Of these three $\xi$ Per shows a non-thermal spectrum and $\alpha$ Cam and $\lambda$ Cep show thermal spectra. As non-thermal radio emission is assumed to be due to synchrotron emission, the detection of a non-thermal radio spectrum in $\xi$ Per strengthens the case that the observed UV line variability observed in this star is caused by a magnetic field.

\subsection{The origin of non-thermal radio emission}
\citet[][see also \citealt{rauw:2002}]{vanloo:2004} were able to successfully reproduce the non-thermal emission of the O5 If star Cyg OB2 No.\ 9, using a model assuming a power-law for the momentum distribution of the relativistic electrons. Reasonable values for the magnetic field that were used are of the order of 10--100 gauss. However, more recent numerical simulations incorporating individual shocks for the acceleration of electrons \citep[e.g.][]{vanloo:2005a,vanloo:2005b} suggest that both a magnetic field and a binary companion are required to explain non-thermal radio emission from massive stars. In these simulations, the synchrotron radiation from single massive stars with magnetic fields is produced relatively close to the star, where shocks occur which accelerate the electrons.
This radiation is absorbed in the stellar wind due to the large free-free opacity. When a star has a massive binary companion, electrons are accelerated in the wind-wind collision region, where the radio emission can escape because of the lower opacity ($\tau_\mathrm{radio} \lesssim 1$).

In the less massive strongly magnetic Bp stars, non-thermal radio emission is observed that is thought to be produced by mildly relativistic electrons which are trapped in the stellar magnetosphere \citep[e.g.][]{drake:1987,leone:1996}. Radio observations of known magnetic massive stars would help to increase our understanding of the relation between magnetic field strength, mass-loss rate and non-thermal radio emission. Unfortunately, the few massive stars that have measured magnetic fields are likely relatively weak radio sources as they have rather weak (order $10^2$ gauss) fields and a low $\dot{M}$ (\object{$\beta$ Cep}, \object{V2052 Oph}, \object{$\zeta$ Cas} and \object{$\tau$ Sco}), or are in a crowded field ($\theta^1$ Ori C).

\subsection{$\xi$ Per, $\lambda$~Cep and $\alpha$~Cam}
$\xi$ Per is a single runaway star \citep[][]{gies:1986}. Runaway stars are stars with high space velocities of $\sim$30-200 km\,s$^{-1}$ and are usually single stars. Therefore, at least for some single massive stars it is possible to have a (mildly) non-thermal radio spectrum. The detection of variability in the radio flux of close binary stars suggest that stellar winds are not as optically thick as generally assumed \citep[][]{blomme:2005}. 
Due to porosity effects, clumping and asphericity of the mass loss due to magnetic fields, it might be possible to observe radio emission from much closer to the central star.
For producing synchrotron emission, a magnetic field is necessary but not sufficient, as relativistic electrons are also required. In the case of $\xi$ Per, and other possibly single massive stars with non-thermal radio spectra, the precise origin of the relativistic electrons needs further investigation. One possible origin in the case of runaway stars would be the bow shock of the star moving through the interstellar material. However, in high resolution radio observations, such as our 3.6 cm observation of $\xi$ Per, such a bow shock would most likely be resolved.

In principle, variability of the radio flux of $\xi$ Per could change the spectral index since the data at different wavelengths are taken at different epochs. However, observed radio variability of massive stars is usually related to the orbit of a massive companion. Such variability is not expected for $\xi$ Per, but as the mechanism responsible for the relativistic electrons is not completely understood we plan future observations to confirm the non-thermal character of the radio spectrum, and check if variability is present.

The runaway stars $\lambda$~Cep and $\alpha$~Cam, which are presumably single \citep{gies:1986}, have a thermal radio spectrum, but it can not be excluded that they have a magnetic field. $\lambda$~Cep has a much denser stellar wind, and the non-thermal emission might all be absorbed. In the case of $\alpha$~Cam, the magnetic field might be too weak to produce observable non-thermal emission, but the lack of non-thermal emission could also be due to the location (closer to the star) or strength of the shocks required to produce the relativistic electrons.

Finally, our results agree with recent results by \citet{fullerton:2006} and \citet{puls:2006} that the mass-loss rates as derived from free-free radio emission are significantly lower than those derived from H$\alpha$ modeling, which is a signature of enhanced clumping in the inner part of stellar wind.

{\acknowledgements We are grateful to J.\ Puls and M.R.\ Mokiem for useful discussions on H$\alpha$ modeling, and to C.\ Stanghellini for discussions on the radio observations of $\xi$ Per from 2004. This research has made use of the Simbad and ADS databases. The Westerbork Synthesis Radio Telescope is operated by ASTRON (the Netherlands Foundation for Research in Astronomy) with support from the Netherlands Foundation for Scientific Research NWO. The Very Large Array is part of the National Radio Astronomy Observatory, which is a facility of the National Science Foundation operated under cooperative agreement by Associated Universities, Inc.}

\bibliographystyle{aa}
\bibliography{../../references}

\end{document}